\documentclass[a4paper]{article}
\usepackage[pdftex]{graphicx}
\usepackage{cite}

\begin{document}

\begin{center} 
	\Large{\textbf{Hidden correlations in indivisible qudits
\newline as a resource for quantum technologies
\newline on examples of superconducting circuits}}

\normalsize{\textbf{Margarita A Man'ko$^1$, Vladimir I Man'ko$^{1,2}$}}\\
\small{$^1$ Lebedev Physical Institute, Leninskii Prospect 53, Moscow 119991, Russia}\\
\small{$^2$ Moscow Institute of Physics and Technology (State University),
Dolgoprudnyi, Moscow Region 141700, Russia}

\tiny{mmanko@sci.lebedev.ru}
\end{center}
\begin{abstract}
We show that the density-matrix states of noncomposite qudit systems
satisfy entropic and information relations like the subadditivity
condition, strong subadditivity condition, and Araki--Lieb
inequality, which characterize hidden quantum correlations of
observables associated with these indivisible systems. We derive
these relations employing a specific map of the entropic
inequalities known for density matrices of multiqudit systems to the
inequalities for density matrices of single-qudit systems. We
present the obtained relations in the form of mathematical
inequalities for arbitrary Hermitian $N$$\times$$N$-matrices. We
consider examples of superconducting qubits and qudits. We discuss
the hidden correlations in single-qudit states as a new resource for
quantum technologies analogous to the known resource in correlations
associated with the entanglement in multiqudit systems.
\end{abstract}

\section{Introduction}

Quantum correlations associated, for example, with the phenomenon of
entanglement~\cite{Schroedinger35} which exists in bipartite and
multipartite systems, like systems of qubits or qudits, provide a
resource for development of quantum technologies~\cite{Chuang-book}.
The properties of correlations can be characterized by different
kinds of entropic and information relations known for both classical
probability distributions and classical observables (see, for
example,~\cite{Holevo-book}) and quantum density matrices of
composite systems, like bipartite and tripartite systems, where
entropic inequalities given as the subadditivity and strong
subadditivity conditions provide information on the presence and
degree of correlations~\cite{Lieb,Ruskai,Petz}.

Among quantum systems studied in connection with possible
applications in quantum technologies, the superconducting circuits
based on Josephson junctions are considered from both theoretical
and experimental points of
view~\cite{Devoret,Martinis,Ustinov,Astafiev,JSLROlga,
Olga-Korean,VI-JRLR,KiktenkoPRA,KiktenkoPLA,GlushkovJRLR,ZeilingerJPA,
ZeilingerPRB,ZeilingerJPCS,ADodonov,FedorovPS}.

Recently, it was pointed out that quantum correlations known for
bipartite and multipartite systems also exist in the systems without
subsystems (indivisible or noncomposite
systems)~\cite{MA-IJQInf,JRLR-Hidden,Entropy,Markovich-JRLR}. We
called such correlations in noncomposite systems the hidden
correlations. For example, analogs of the entropic and information
inequalities like the subadditivity condition, the strong
subadditivity condition, and the Araki--Lieb inequality~\cite{Araki}
expressed in the form of matrix relations for density matrices of
single qudit systems exist for indivisible systems as well.

The aim of this paper is to review the approach developed in
\cite{JRLR-Hidden,Entropy} for obtaining new entropic inequalities
for systems without subsystems and extend this approach for
obtaining new entropic inequalities not only for the state density
matrices but also for Hermitian matrices of observables. The idea of
the approach is based on the fact that any Hermitian matrix can be
mapped on the nonnegative matrix with unit trace. The corresponding
tool was applied in \cite{OcastaJPA} to obtain the relation of
entropy $S$ and energy $E$ of qudit systems, $E+S\leq\ln Z(-1)$,
where $Z(\beta)$ is the system partition function, with $\beta$
being the inverse temperature.

The approach formulated provides the possibility to extend all
density-matrix inequalities known for multipartite systems to
density-matrix inequalities for indivisible systems as well as to
all observables. Here, we concentrate on particular examples of
superconducting qudits and finite-dimensional systems.

This paper is organized as follows.

In section~2, we review the properties of superconducting circuit
states modeled by a parametric quantum oscillator with vibrating
voltage and current. In section~3, we present a new inequality for
the classical single qudit system in the form of subadditivity
condition. In section~4, we discuss the approach to study Hermitian
matrices using a map of the matrices on nonnegative matrices to
derive new inequalities for Hermitian matrices. In section~5, we
consider examples of inequalities for classical probabilities and
classical observables. In section~6, we obtain new inequalities for
quantum observables on the example of artificial atom realized by a
superconducting qudit with $j=3/2$. In section~7, we give the
conclusions and prospectives.

\section{Superconducting qubits}

Superconducting devices based on application of Josephson junctions
are discussed~\cite{Devoret,Martinis,Ustinov,Astafiev} as possible
technical instruments for developing new technologies, for example,
quantum computing. The idea of such applications is related to the
fact that the Josephson junction realizes a model of the electric
circuit with inductance $L$ and capacitance $C$~\cite{JSLROlga},
i.e., the current and voltage in devices with Josephson junctions
vibrate as the momentum and position of a mechanical oscillator does
(see, for example,~\cite{LightYearJRLR}). The frequency of
vibrations $\omega$ in the electric circuit is determined by a
factor proportional to $(LC)^{-1/2}$, and for high frequencies and
low temperatures, such that $\hbar\omega\geq T$, the oscillator
behaves as a quantum oscillator called the superconducting circuit,
where the current and voltage satisfy the Heisenberg uncertainty
relation~\cite{Heisenberg27}, the Schr\"odinger--Robertson
uncertainty relation~\cite{Schroedinger30,Robertson29}, and the
purity-dependent uncertainty relation~\cite{183}.

If the circuit parameters $L$ and $C$ are constant, the stationary
states of the quantum oscillator correspond to the energy levels
$E_n$ $(n=0,1,2,\ldots)$. If the oscillator excitations are such
that only finite number of levels $N$ is involved, the stationary
states of the superconducting circuit are identified with the qudit
states of $j=(N-1)/2$. For example, if only two oscillator energy
levels are involved, the superconducting circuit realizes a
superconducting qubit and, in this case, a set of Josephson
junctions models a multiqubit system, which can be employed to
provide quantum information devices.

To realize the dynamical (nonstationary) Casimir effect, in
\cite{JSLROlga,Olga-Korean,VI-JRLR} it was suggested to use the Josephson
junction with time-dependent parameters [inductance $L(t)$ and
capacitance $C(t)$], which is an analog of the oscillator with
time-dependent frequency $\omega(t)$. The current and voltage in
parametric superconducting circuits are generated due to temporal
variations in the Josephson-junction parameters analogously to the
photon in squeezed-states generation in resonators with vibrating
boundaries due to the dynamical Casimir effect. The photons created
due to the dynamical Casimir effect were registered in the devices
where the Josephson junctions are employed~\cite{Johansson-PRL}.

The oscillator with time-dependent frequency is described by the Hamiltonian
(we use dimensionless variables $\hbar=m=\omega(0)=1$)
\begin{equation}\label{1}
\hat H=\frac{\hat p^2}{2}+\omega^2(t)\frac{\hat q^2}{2}\,.
\end{equation}
In \cite{Malkin-PLA70}, it was found that the parametric oscillator
has linear in the position and momentum integrals of motion called
dynamical invariants (see, for example,
\cite{183,Guerrero150,SuslovZ}) of the form
\begin{equation}\label{2}
\hat a(t)=-\frac{i}{\sqrt2}\left(\dot{\cal E}(t)\hat q-{\cal E}(t)\hat
p\right),
\end{equation}
where the complex function ${\cal E}(t)$ satisfies the classical
equation of motion for a parametric oscillator $\ddot{\cal
E}(t)+\omega^2(t){\cal E}(t)=0$ under the initial conditions ${\cal
E}(t)=1$,  $\dot{\cal E}(0)=i$, and $\hat
a(0)=\displaystyle{\frac{1}{\sqrt2}}\left(\hat q+i\hat p\right)$.

The linear invariants $\hat a(t)$ and $\hat a^\dagger (t)$ satisfy
the commutation relation $\left[\hat a(t),\hat a^\dagger
(t)\right]=1$.

For $\omega(t)=1$, the integral of motion $\hat a(t)=e^{it}\hat
a(0)$.

An analog of the ground state of the stationary oscillator
(superconducting circuit), being the squeezed state, has the wave
function
\begin{equation}\label{4}
\psi_0(x,t)=\left(\pi{\cal E}^2(t)\right)^{-1/4}\exp\frac{i\dot{\cal
E}(t)x^2}{2{\cal E}(t)}\,.
\end{equation}
Dispersions of the position $\sigma_{xx}(t)$ and momentum $\sigma_{pp}(t)$
(current and voltage in a superconducting circuit) are determined by the
function ${\cal E}(t)$ as follows:
\begin{equation}\label{5}
\sigma_{xx}(t)=\frac{|{\cal E}(t)|^2}{2}\,,\qquad
\sigma_{pp}(t)=\frac{|\dot{\cal E}(t)|^2}{2}\,.
\end{equation}
The correlation coefficient $r(t)$ of the current and voltage
$r(t)=\displaystyle{\frac{\sigma_{xp}(t)}{\sqrt{\sigma_{xx}(t)\sigma_{pp}(t)}}}$
is given by the bound in the Schr\"odinger--Robertson uncertainty relation
\begin{equation}\label{6}
\sigma_{xx}(t)\sigma_{pp}(t)\geq\left[4\left(1-r^2(t)\right)\right]^{-1},
\end{equation}
which provides the equality
\begin{equation}\label{7}
|{\cal E}(t)\dot{\cal E}(t)|^2= \left[4-r^2(t)\right]^{-1}.
\end{equation}

The eigenfunctions of time-dependent invariants $\hat n(t)=\hat
a^\dagger(t)\hat a(t)$ are analogs of stationary states of superconducting
circuits with time-dependent parameters $L$ and $C$. The eigenvalues of the
integral of motion $\hat n(t)$ do not depend on time and take the values
$n=0,1,2, \ldots,\infty$.

For finite number of excited states $\mid n\rangle$ $(n=0,1,2,
\ldots,N)$ of a parametric superconducting circuit, the states can
be considered as an approximation of qudit states with $j=(N-1)/2$.
These states can be used analogously to the states of stationary
superconducting circuits in quantum technologies. For example, if
$j=1/2$, a parametric superconducting qubit can be realized.

The symplectic tomogram (probability
distribution)~\cite{ManciniPLA96} of the squeezed vacuum
state~(\ref{4}) of the superconducting circuit
\begin{equation} \label{8}
w(X,t\mid\mu,\nu)=\frac{1}{2\pi|\nu|}\left|\int\psi_0(y,t)\exp\left(
\frac{i\mu}{2\nu}y^2-\frac{iX}{\nu}y\right)dy\right|^2
\end{equation}
reads
\begin{equation} \label{9}
w(X,t\mid\mu,\nu)=\frac{1}{\sqrt{2\pi\sigma_{xx}(t)}}
\exp\left(-\frac{X^2}{2\sigma_{xx}(t)}\right),
\end{equation}
where the dispersion $\sigma_{xx}(t)$ of the homodyne quadrature $X$
is
\begin{equation} \label{10}
\sigma_{xx}(t)=\mu^2\frac{|{\cal E}(t)|^2}{2}+\nu^2\frac{|\dot{\cal
E}(t)|^2}{2}+\frac12\mu\nu\left(|{\cal E}(t)\dot{\cal
E}(t)|^2-1\right),
\end{equation}
with $\mu$ and $\nu$ being the real parameters.

The optical tomogram $w(X,t\mid\theta)$ of squeezed vacuum state of
the superconducting circuit has the form~(\ref{9}) with the
dispersion $\sigma_{xx}(t)$ of the homodyne quadrature $X$ dependent
on the local oscillator phase as
\begin{equation} \label{11}
\sigma_{xx}(t)=\frac{\cos^2\theta|{\cal
E}(t)|^2}{2}+\frac{\sin^2\theta|\dot{\cal E}(t)|^2}{2}+\frac{\sin
2\theta}{2}\left(|{\cal E}(t)\dot{\cal E}(t)|^2-1\right).
\end{equation}

The squeezed coherent state of the parametric superconducting
circuit $\mid\alpha,t\rangle$ is such that $\hat
a(t)\mid\alpha,t\rangle=\alpha\mid\alpha,t\rangle$ has the
symplectic tomogram in the form of normal distribution with
dispersion $\sigma_{xx}(t)$ given by (\ref{10}) and the mean value
of the homodyne quadrature $X$
\begin{equation} \label{12}
\langle X(t)\rangle=\mu\frac{{\cal E}^\star(t)\alpha+{\cal
E}(t)\alpha^\star}{\sqrt 2}+\nu\frac{\dot{\cal E}^\star(t)\alpha+\dot{\cal
E}(t)\alpha^\star}{\sqrt 2}\,.
\end{equation}
Thus, the optical tomogram of the squeezed coherent state of the
superconducting circuit reads
\begin{equation} \label{13}
w_\alpha(X,t\mid\theta)=\frac{1}{\sqrt{2\pi\sigma_{xx}(t)}}
\exp\left(-\frac{X-\bar X^2_\alpha(t)}{2\sigma_{xx}(t)}\right),
\end{equation}
where $\sigma_{xx}(t)$ is given by (\ref{11}), and the mean value of the
homodyne quadrature is
\begin{equation} \label{14}
\bar X_\alpha(t)=\sqrt2\,\mbox{Re}\left[\alpha\left({\cal E}^\star(t)
\cos\theta+\dot{\cal E}^\star(t)\sin\theta\right)\right].
\end{equation}

One can check that tomogram~(\ref{13}) satisfies the entropic inequality valid
for an arbitrary optical tomogram~\cite{FP-MA}
\begin{equation} \label{15}
-\int w(X,t\mid\theta)\ln w(X,t\mid\theta)\,dX-\int w(X,t\mid\theta+\pi/2)\ln
w(X,t\mid\theta+\pi/2)\,dX\geq\ln (\pi e).
\end{equation}
For example, the state $\mid 1,t\rangle=\hat a^\dagger(t)\mid 0,t\rangle$ with
the wave function
\begin{equation}\label{16}
\psi_1(x,t)=-\frac{i}{\sqrt2}\left({\cal E}^\star(t)\hat p-\dot{\cal
E}^\star(t)x\right)\psi_0(x,t),
\end{equation}
which is the second component in the superconducting qubit state with the
first component given by (\ref{4}), has the tomogram
\begin{equation} \label{17}
w_1(X,t\mid\theta)=\pi^{-1/2}\big(2\sigma_{xx}(t)\big)^{-3/2}X^2
\exp\left(-\frac{X^2}{2\sigma_{xx}(t)}\right),
\end{equation}
where the dispersion $\sigma_{xx}(t)$ is given by (\ref{11}).
Tomogram~(\ref{17}) satisfies the entropic inequality~(\ref{15}). Optical
tomograms of the superconducting circuit states $\mid
n,t\rangle=\displaystyle{\frac{\hat a^\dagger(t,n)}{\sqrt{n!}}}\mid
0,t\rangle$ are the probability distributions
\begin{equation} \label{18}
w_n(X,t\mid\theta)=w_0(X,t\mid\theta)\frac{1}{2^nn!}H_n^2\left(\frac{X}{\sqrt{2\sigma_{xx}(t)}}\right),
\end{equation}
where $H_n(y)$ are Hermite polynomials. The dependence of the
tomogram on the local oscillator phase $\theta$ is given by the
dependence~(\ref{11}) of the dispersion $\sigma_{xx}(t)$ on this
parameter.

If the states $\mid n,t\rangle$ with $n=0,1,2,\ldots,N=2j+1$ are excited, the
parametric superconducting circuit can be interpreted as an artificial atom
with $N$ levels or qudit with $j=(N-1)/2$.

\section{Information inequalities for single qudit states}

In this section, we discuss the entropic and information relations,
such as equalities and inequalities, as well as quantum correlations
of qudit observables for single qudits. Qudit can be realized either
by a spin-$j$ particle or the $N$-level atom. These systems can also
be considered as artificial atoms realized by superconducting
circuits. In this consideration, we
follow~\cite{Entropy,NuovoCim16,VovaPS-150,JRLR-Deformed} where a
map of integers $1,2,\ldots,N=mn$ onto pairs of integers $(jk)$,
$j=1,2,\ldots,m$ and $k=1,2,\ldots,n$ or, in the case of
$N=n_1n_2n_3$, on triples of integers $(jkl)$, $j=1,2,\ldots,n_1$,
$k=1,2,\ldots,n_2$ and $l=1,2,\ldots,n_3$ was used. This map
provides a possibility to apply the relations like entropic
inequalities (known for bipartite and tripartite quantum systems) to
indivisible systems like a superconducting-circuit qudit.

We demonstrate the inequalities for the probability distribution
$P_s=(P_1,P_2,\ldots,P_N)$, where the even number $N=n_1n_2$,
$n_1=2$, and $n_2=N/2$. Then we label the integers $s$, where
$s=1,2,\ldots,N$, by the pairs of integers $(jk)$, where $j=1,2$ and
$k=1,2,\ldots,N/2$. We obtain the same set of numbers $P_s\equiv
P_{s(jk)}\equiv P_{jk}$. The inequality, known as the subadditivity
condition for the joint probability distribution $P_{jk}$, for $P_s$
reads
\begin{eqnarray} \label{19}
-\sum_{s=1}^NP_s\ln P_s\leq
-\left(\sum_{s=1}^{N/2}P_s\right)\ln\left(\sum_{s=1}^{N/2}P_s\right)
-\left(\sum_{s=N/2}^{N}P_s\right)\ln\left(\sum_{s=N/2}^{N}P_s\right)\nonumber\\
-\sum_{k=1}^{N/2}(P_k+P_{k+N/2})\ln (P_k+P_{k+N/2}).
\end{eqnarray}

\section{Hermitian matrix inequalities}
Given Hermitian $N$$\times$$N$-matrix $h$, where $N=nm$. Let the
eigenvalues of the matrix $h$ have values with a minimal eigenvalue
$h_0$. The Hermitian $N$$\times$$N$-matrix $\rho(x)$, where $x>h_0$
given in the block form with $n$$\times$$n$-blocks
\begin{equation} \label{20}
\rho(x)=(Nx+\mbox{Tr}\,h)^{-1}(h^{jk}+x1_n\delta^{jk}),\qquad
j,k=1,2,\ldots,m,
\end{equation}
is the nonnegative matrix with Tr$\,\rho(x)=1$, where $1_n$ is the
identity $n$$\times$$n$-matrix. For the density matrix $h=\rho$, the
map $\rho\to\rho(x)$ given by (\ref{20}) reflects the fact that the
noise contribution is taken into account.

One can check that two Hermitian matrices, namely, $m$$\times$$m$-matrix
$\rho(1,x)$ with matrix elements
\begin{equation} \label{21}
\rho_{jk}(1,x)=(Nx+\mbox{Tr}\,h)^{-1}\left[\mbox{Tr}(h^{jk})+nx\delta^{jk}\right]
\end{equation}
and $n$$\times$$n$-matrix $\rho(2,x)$ of the form
\begin{equation} \label{22}
\rho(2,x)=(Nx+\mbox{Tr}\,h)^{-1}\left(mx1_n+\sum_{k=1}^mh^{kk}\right)
\end{equation}
are nonnegative matrices and Tr$\,\rho(1,x)=\mbox{Tr}\,\rho(2,x)=1$.
The matrices satisfy the entropic inequality for the mutual
information $I(x)$ of the form
\begin{equation} \label{23}
I(x)=\mbox{Tr}\,\rho(x)\ln\rho(x)-\mbox{Tr}\,\rho(1,x)\ln\rho(1,x)
-\mbox{Tr}\,\rho(2,x)\ln\rho(2,x)\geq 0.
\end{equation}
If $h_0\geq 0$, one can assume $x=0$. In this case,
inequality~(\ref{23}) for an arbitrary Hermitian
$N$$\times$$N$-matrix $h$ reads
\begin{eqnarray} \label{24}
I(x)=\mbox{Tr}\left\{h\ln(h/\mbox{Tr}\,h)\right\}-\mbox{Tr}\left\{\left[\mbox{Tr}\,h^{jk}\right]\ln\left[
\left(\mbox{Tr}\,h^{jk}\right)/\,\mbox{Tr}\,h\right]\right\}\nonumber\\
-\mbox{Tr}\left\{\left[\sum_{k=1}^mh^{kk}\right]\ln\left[
\left(\sum_{k=1}^mh^{kk}\right)/\,\mbox{Tr}\,h\right]\right\}\geq 0.
\end{eqnarray}

Now we present this inequality on an example of the 3$\times$3-matrix
$h=\left(\begin{array}{ccc}
h_{11}&h_{12}&h_{13}\\
h_{21}&h_{22}&h_{23}\\
h_{31}&h_{32}&h_{33}\end{array}\right)$. Since $N=3$, we may assume this
matrix as $\tilde h=\left(\begin{array}{cc}
h&0\\
0&0\end{array}\right)$. Then taking in the previous formulas $m=n=2$, we
obtain the 4$\times$4-matrix $\rho(x)$~(\ref{21}) as follows:
\begin{equation} \label{25}
\rho(x)=(3x+h_{11}+h_{22}+h_{33})^{-1}\left(\begin{array}{cc}
h^{11}+x1_2&h^{12}\\
h^{21}&h^{22}+x{\cal P}_1\end{array}\right),
\end{equation}
where ${\cal P}_1=\left(\begin{array}{cc}
1&0\\
0&0\end{array}\right)$ and 2$\times$2-blocks $h^{jk}$ $(j,k=1,2)$ read
$$
h^{11}=\left(\begin{array}{cc}
h_{11}&h_{12}\\
h_{21}&h_{22}\end{array}\right),\quad h^{12}=\left(\begin{array}{cc}
h_{13}&0\\
0&0\end{array}\right),\quad h^{21}=\left(\begin{array}{cc}
h_{31}&0\\
0&0\end{array}\right),\quad h^{22}=\left(\begin{array}{cc}
h_{33}&0\\
0&0\end{array}\right).$$ Then the 2$\times$2-matrix $\rho(1,x)$ has the
form
\begin{equation} \label{26}
\rho(1,x)=(3x+h_{11}+h_{22}+h_{33})^{-1}\left(\begin{array}{cc}
h_{11}+h_{22}+2x&h_{13}\\
h_{31}&h_{33}+x\end{array}\right),
\end{equation}
and the 2$\times$2-matrix $\rho(2,x)$ reads
\begin{equation} \label{27}
\rho(2,x)=(3x+h_{11}+h_{22}+h_{33})^{-1}\left(\begin{array}{cc}
h_{11}+h_{33}+2x&h_{12}\\
h_{21}&h_{22}+x\end{array}\right).
\end{equation}
Now we are in the position to present the entropic inequality which
is satisfied by the matrix elements of the 3$\times$3-matrix $h$ for
$x>\delta_0$; it is

\noindent
\scalebox{0.865}{
\parbox{\textwidth}{
\begin{eqnarray} \label{28}
     I=\mbox{Tr}\left\{\left(\begin{array}{ccc}
h_{11}+x&h_{12}&h_{13}\\
h_{21}&h_{22}+x&h_{23}\\
h_{31}&h_{32}&h_{33}+x\end{array}\right) \ln\left[\left(\begin{array}{ccc}
h_{11}+x&h_{12}&h_{13}\\
h_{21}&h_{22}+x&h_{23}\\
h_{31}&h_{32}&h_{33}+x\end{array}\right)(3x+h_{11}+h_{22}+h_{33})^{-1}\right]\right\}
   \nonumber\\
-\mbox{Tr}\left\{\left(\begin{array}{cc}
h_{11}+h_{22}+2x&h_{13}\\
h_{31}&h_{33}+x\end{array}\right) \ln \left[\left(\begin{array}{cc}
h_{11}+h_{22}+2x&h_{13}\\
h_{31}&h_{33}+x\end{array}\right)(3x+h_{11}+h_{22}+h_{33})^{-1}\right]\right\} \nonumber\\
-\mbox{Tr}\left\{\left(\begin{array}{cc}
h_{11}+h_{33}+2x&h_{13}\\
h_{21}&h_{22}+x\end{array}\right) \ln\left[\left(\begin{array}{cc}
h_{11}+h_{33}+2x&h_{13}\\
h_{21}&h_{22}+x\end{array}\right)(3x+h_{11}+h_{22}+h_{33})^{-1}\right]\right\}\geq
0.\nonumber\\
\end{eqnarray}}}\par
If the matrix $h$ is nonnegative Hermitian matrix, one has
inequality~(\ref{28}), where $x=0$; this inequality reads
\begin{eqnarray} \label{29}
I=\mbox{Tr}\left\{\left(\begin{array}{ccc}
h_{11}&h_{12}&h_{13}\\
h_{21}&h_{22}&h_{23}\\
h_{31}&h_{32}&h_{33}\end{array}\right) \ln\left[\left(\begin{array}{ccc}
h_{11}&h_{12}&h_{13}\\
h_{21}&h_{22}&h_{23}\\
h_{31}&h_{32}&h_{33}\end{array}\right)(h_{11}+h_{22}+h_{33})^{-1}\right]\right\}\nonumber\\
-\mbox{Tr}\left\{\left(\begin{array}{cc}
h_{11}+h_{22}&h_{13}\\
h_{31}&h_{33}\end{array}\right) \ln \left[\left(\begin{array}{cc}
h_{11}+h_{22}&h_{13}\\
h_{31}&h_{33}\end{array}\right)(h_{11}+h_{22}+h_{33})^{-1}\right]\right\}\nonumber\\
-\mbox{Tr}\left\{\left(\begin{array}{cc}
h_{11}+h_{33}&h_{13}\\
h_{21}&h_{22}\end{array}\right) \ln\left[\left(\begin{array}{cc}
h_{11}+h_{33}&h_{13}\\
h_{21}&h_{22}\end{array}\right)(h_{11}+h_{22}+h_{33})^{-1}\right]\right\}\geq
0.
\end{eqnarray}
If in inequality~(\ref{29}) $\mbox{Tr}\,h=1$ and $h\geq 0$, one
arrives to the inequality obtained for the qutrit density matrix in
\cite{Vova-portrait}. The discussed inequalities describe relations
for the density matrix of qutrit corresponding to an artificial
three-level atom.

\section{Classical observables as ``classical states''}

Given random variables $j=1,2,\ldots,N$ and the probability
distribution $P_j$, which we call ``the classical state.'' One has
$\sum_{j=1}^NP_j=1$. Also given a real function $f_j$ of random
variables, one can interpret the function $f_j$ as an observable.
The means and higher moments are
\begin{equation} \label{30}
\langle f\rangle=\sum_{j=1}^Nf_jP_j,\qquad\langle f^k\rangle=\sum_{j=1}^Nf_j^kP_j.
\end{equation}
One can construct a new function $F_j=(f_j+x_j)n$, where
$x>|f_{\tilde j}|$. The number $f_{\tilde j}$ is a minimum value of
the observable $f_j$, and it can be negative. The normalization
factor $n=\left[\left(\sum_{j=1}^Nf_j\right)+Nx\right]^{-1}$ and the
function $F_j$ is nonnegative and normalized, $\sum_{j=1}^NF_j=1$.
We can consider the function $F_j$ as a probability distribution,
which we associate with the observable $f_j$, and this fact provides
the possibility to extend some relations known for the probability
distributions and apply these relations to the observables. For
example, we can introduce the Shannon entropy~\cite{Shannon} for the
observable $f_j$ as follows:
\begin{equation} \label{31}
S_f(x)=-\sum_{j=1}^N\left[(f_j+x)n\ln (f_j+x)n\right] \geq 0.
\end{equation}
Also we can associated the Tsallis $q$-entropy~\cite{Tsallis} with
the observable $f_j$ as
\begin{equation} \label{32}
T_f(x)=\frac{1}{1-q}\left\{\sum_{j=1}^N\left[(f_j+x)n\right]^q-1\right\}.
\end{equation}
Then we can rewrite the known inequality for relative
entropy~\cite{Chuang-book} associated with two probability
distributions for two observables $f_j^{(1)}$ and $f_j^{(2)}$; it
reads
\begin{equation} \label{33}
\sum_{j=1}^N\left[\left(f_j^{(1)}+x_1\right)n_1\ln\left(f_j^{(1)}+x_1\right)n_1\right]-
\left[\left(f_j^{(1)}+x_1\right)n_1\ln\left(f_j^{(2)}+x_2\right)n_2\right]\geq
0.
\end{equation}
Here, $x_1$ and $x_2$ are moduli of minimum nonpositive values of
observables $f_j^{(1)}$ and $f_j^{(2)}$, respectively. In
particular, one can write the inequality for an arbitrary observable
$f_j$ and the probability distribution $P_j$
\begin{equation} \label{34}
\sum_{j=1}^N\big[P_j\ln P_j-P_j\ln[(f_j+x)n]\big] \geq 0.
\end{equation}
Analogously, we can write inequalities for observables in view of
Tsallis entropy.

The subadditivity condition is valid for an arbitrary observable
$f_j$ as well; for a single system with $N=4$, we provide the
condition in an explicit form. Thus, given an observable $f_j$,
i.e., four numbers $f_1$,  $f_2$,  $f_3$, and $f_4$. Assume the
number $f_4$ to be negative, i.e., $f_4=-|f_4|$ and take the
variable $x>|f_4|$. Now we introduce the function $F_j(x)$ taking
four values
\begin{equation} \label{35}
F_1(x)=(f_1+x)n_4,~~F_2(x)=(f_2+x)n_4,~~F_3(x)=(f_1+x)n_4,~~
F_4(x)=(f_4+x)n_4,
\end{equation}
where $n_4=(f_1+f_2+f_3+f_4+4x)^{-1}$. Then we have the
nonnegativity condition for relative entropy associated with the
probability distribution $P_j$ and observable $f_j$ of the form
\begin{eqnarray} \label{36}
&&~~P_1\ln\big\{P_1[(f_1+x)n_4]^{-1}\big\}+P_2\ln\big\{P_2[(f_2+x)n_4]^{-1}\big\}\nonumber\\
&&+P_3\ln\big\{P_3[(f_3+x)n_4]^{-1}\big\}+P_4\ln\big\{P_4[(f_4+x)n_4]^{-1}\big\}\geq
0.
\end{eqnarray}

The other known inequality is the subadditivity condition for
entropy associated with the observable $f_j$ and two other
entropies, which are analogs of ``subsystem'' states described by
two probability distributions
$${\cal P}_1(x)=(f_1+f_2+2x)n_4,\qquad {\cal
P}_2(x)=(f_3+f_4+2x)n_4$$ and $$\Pi_1(x)=(f_1+f_3+2x)n_4,\qquad
\Pi_2(x)=(f_2+f_4+2x)n_4.$$
The inequality
\begin{equation} \label{37}
S_1(x)+S_2(x)\geq S(x),
\end{equation}
where
$$
S_1(x)=-\sum_{k=1}^2{\cal P}_k(x)\ln {\cal P}_k(x),\quad
S_2(x)=-\sum_{k=1}^2\Pi_k(x)\ln\Pi_k(x),\quad
S(x)=-\sum_{k=1}^4F_k(x)\ln F_k(x),
$$
reads
\begin{eqnarray} \label{38}
&&-(f_1+f_2+2x)\ln[(f_1+f_2+2x)n_4]-(f_3+f_4+2x)\ln[(f_3+f_4+2x)n_4]\nonumber\\
&&-(f_1+f_3+2x)\ln[(f_1+f_3+2x)n_4]-(f_2+f_4+2x)\ln[(f_2+f_4+2x)n_4]\nonumber\\
&\geq &-(f_1+x)\ln[(f_1+x)n_4]-(f_2+x)\ln[(f_2+x)n_4]\nonumber\\
&&-(f_3+x)\ln[(f_3+x)n_4]-(f_4+x)\ln[(f_4+x)n_4].
\end{eqnarray}
Thus, the values of an arbitrary classical observable $f_j$ satisfy
the inequality, which is an analog of the subadditivity condition
for entropies of bipartite classical systems. Formally,
inequality~(\ref{37}) is valid for arbitrary real numbers $f_j$ for
large numbers $x$.

\section{Quantum inequalities}
Hidden correlations described by classical probability distributions
and characterized by entropic inequalities take place in quantum
systems. For example, the four-level artificial atom realized in the
superconducting circuit as a qudit with $j=3/2$ is described by the
density 4$\times$4-matrix $\rho=
\left(\begin{array}{cccc}
\rho_{11}&\rho_{12}&\rho_{13}&\rho_{14}\\
\rho_{21}&\rho_{22}&\rho_{23}&\rho_{24}\\
\rho_{31}&\rho_{32}&\rho_{33}&\rho_{34}\\
\rho_{14}&\rho_{24}&\rho_{34}&\rho_{44}
\end{array}\right)$. One can check that this Hermitian matrix
satisfies the entropic subadditivity condition in spite of the fact
that the system under consideration is not bipartite; the inequality
reads
\begin{equation} \label{39}
-\mbox{Tr}\,(\rho\ln\rho)\leq -\mbox{Tr}\,[\rho(1)\ln\rho(1)]
-\mbox{Tr}\,[\rho(2)\ln\rho(2)],
\end{equation}
where $\rho(1)=\left(\begin{array}{cc}
\rho_{11}+\rho_{22}&\rho_{13}+\rho_{34}\\
\rho_{31}+\rho_{41}&\rho_{33}+\rho_{44}
\end{array}\right)$ and $\rho(2)=\left(\begin{array}{cc}
\rho_{11}+\rho_{33}&\rho_{12}+\rho_{34}\\
\rho_{21}+\rho_{42}&\rho_{22}+\rho_{44}
\end{array}\right)$. An observable for the artificial atom is
described by the Hermitian 4$\times$4-matrix $f=
\left(\begin{array}{cccc}
f_{11}&f_{12}&f_{13}&f_{14}\\
f_{21}&f_{22}&f_{23}&f_{24}\\
f_{31}&f_{32}&f_{33}&f_{34}\\
f_{14}&f_{24}&f_{34}&f_{44}
\end{array}\right)$. The matrix $f(1)=\left(\begin{array}{cc}
f_{11}+f_{22}&f_{13}+f_{24}\\
f_{31}+f_{42}&f_{33}+f_{44}
\end{array}\right)$ is the Hermitian 4$\times$4-matrix.
An analog of the relative entropy nonnegativity can be given through
matrices $\rho(1)$ and $f(1)$ as the following inequality:
\begin{eqnarray} \label{40}
&&\mbox{Tr}\left[\left(\begin{array}{cc}
\rho_{11}+\rho_{22}&\rho_{13}+\rho_{34}\\
\rho_{31}+\rho_{41}&\rho_{33}+\rho_{44}
\end{array}\right)\ln\left(\begin{array}{cc}
\rho_{11}+\rho_{22}&\rho_{13}+\rho_{34}\\
\rho_{31}+\rho_{41}&\rho_{33}+\rho_{44}
\end{array}\right)-\left(\begin{array}{cc}
\rho_{11}+\rho_{22}&\rho_{13}+\rho_{34}\\
\rho_{31}+\rho_{41}&\rho_{33}+\rho_{44}
\end{array}\right)\right]\nonumber\\
&\times&\ln\left\{\left(f_{11}+f_{22}+f_{33}+f_{44}+
2x\right)^{-1}\left(\begin{array}{cc}
f_{11}+f_{22}+x&f_{13}+f_{24}\\
f_{31}+f_{42}&f_{33}+f_{44}+x
\end{array}\right)\right\}\geq 0.
\end{eqnarray}
This new inequality is valid for large $x$  for an arbitrary state
of the superconducting circuit with excited four energy levels and
the observable.

A new inequality can also be written for qudit tomograms and
tomographic symbols of observables. For example, for the case of
four-level artificial atom realized by the superconducting circuit,
the tomogram of observable $f$ reads
\begin{equation}\label{41}
w_f(m,\vec n)=(ufu^\dagger)_{mm},
\end{equation}
where $m=1,2,3,4$ and $u$ is the unitary 4$\times$4-matrix depending
on the unit vector $\vec n$. In this case, one has four real
numbers: $w_f(1,\vec n)$, $w_f(2,\vec n)$, $w_f(3,\vec n)$, and
$w_f(4,\vec n)$. Then a new quantum inequality for observable $f$
and the state tomogram with the density matrix $\rho$ can be written
in the form of tomographic relative entropy nonnegativity as
follows:
\begin{equation}\label{42}
\sum_{m=1}^4\left\{w_\rho(m,\vec n)\ln w_\rho(m,\vec n)-w_\rho(m,\vec n)
\ln\left(\big[w_f(m,\vec n)+x\big]\left[\sum_{m'=1}^4\big[w_f(m',\vec
n)+x\big]\right]^{-1}\right)\right\}\geq 0.
\end{equation}
Thus, we obtained some new quantum inequalities corresponding to
quantum correlations available in superconducting qudits taking into
account the properties of observables.

\section{Conclusions}
To conclude, we point out the main results of our study.

We presented a review of the approach for extending the entropic
inequalities known for multipartite systems, both classical and
quantum, for example, the nonnegativity condition of relative
entropy to obtain new inequalities for indivisible systems,
including new inequalities for physical observables.

We discussed the example of superconducting circuit and entropic
inequalities for an artificial atom with four energy levels and its
parametric analog.

We derive a new inequality for the tomographic symbol of an
observable of the four-level atom related to quantum correlations in
the system.

The new inequalities obtained can be checked experimentally
following the procedure considered in
\cite{GlushkovJRLR}, where testing entropic inequalities in
superconducting qudits were discussed.

The new entropic and information inequalities reflect the presence
of quantum correlations in systems like superconducting single
qudits. These correlations can be used as a resource for quantum
technologies analogously to the correlations related to the
entanglement in multiqubit systems and studied as such a resource.
In the future publication, we extend our study to the other systems.


\end{document}